\documentclass[%twocolumn,%
11pt,tightenlines,%
nofootinbib,preprintnumbers,prd]{revtex4}
\usepackage{graphicx}
\newcommand{\be}{\begin{equation}}
\newcommand{\ee}{\end{equation}}
\newcommand{\bea}{\begin{eqnarray}}
\newcommand{\eea}{\end{eqnarray}}
\newcommand{\nn}{\nonumber}

\begin{document}
\preprint{YITP-SB-02-34}
\preprint{NSF-ITP-02-47}
\preprint{TUW-02-13}
\title{Clash of discrete symmetries
for the supersymmetric kink on a circle}

\author{Alfred Scharff Goldhaber}\email{goldhab@insti.physics.sunysb.edu}
\affiliation{C.N.Yang Institute for Theoretical
Physics, State University of New York, Stony Brook, NY 11794-3840 }
\affiliation{Kavli Institute for Theoretical Physics,
University of California, Santa Barbara, CA 93106}

\author{Anton Rebhan}\email{rebhana@hep.itp.tuwien.ac.at}
\affiliation{Institut f\"ur Theoretische Physik, Technische
Universit\"at Wien, %\\Wiedner Haupstr.~8-10, 
A-1040 Vienna, Austria }
\affiliation{Kavli Institute for Theoretical Physics,
University of California, Santa Barbara, CA 93106}

\author{Peter van
Nieuwenhuizen}\email{vannieu@insti.physics.sunysb.edu}
\affiliation{C.N.Yang Institute for Theoretical
Physics, State University of New York, Stony Brook, NY 11794-3840 }

\author{Robert Wimmer}\email{rwimmer@hep.itp.tuwien.ac.at}
\affiliation{Institut f\"ur Theoretische Physik, Technische
Universit\"at Wien, %\\Wiedner Haupstr.~8-10, 
A-1040 Vienna, Austria}

\begin{abstract}
We consider the $N$=1 supersymmetric kink on a circle,
i.e., on a finite interval with boundary or transition conditions which are
locally invisible. 
For Majorana fermions, the single-particle Dirac Hamiltonian as a
differential operator obeys simultaneously the three discrete symmetries
of charge conjugation, parity, and time reversal. 
However, no single locally invisible transition condition can
satisfy all three. 
When calculating sums over zero-point energies by mode number
regularization, 
this gives a new rationale for a
previous suggestion that 
one has to average over
different choices of boundary conditions,
such that for the combined set all three symmetries are obeyed.
In particular it is shown that for twisted periodic or twisted
antiperiodic boundary conditions separately both parity and
time reversal are
violated in the kink sector, as manifested by a delocalized momentum
that cancels only in the average.
\end{abstract}

\maketitle

\section{Introduction}

Subtleties in the application of the discrete symmetries  charge
conjugation ${\cal C}$,
parity ${\cal P}$, and  time reversal ${\cal T}$ to Majorana fermions
%(in particular,
%Majorana neutrinos)  
have long been a topic of interest \cite{RYT,BK}.  
Past discussions
generally have dealt with local processes and properties,  but the main
aim of the present
work is to study an anomalous  global behavior of these discrete
symmetries in a model with
a topological structure. For this we consider the simplest possible
system: the
supersymmetric (susy) kink with what would seem to be natural  boundary
conditions.  

Some time ago the concept of locally  
invisible boundary conditions was introduced
\cite{misha,glv}: for a two component Majorana fermion in a kink
background  in a box of
length $L$, the twisted periodic (TP) and  twisted antiperiodic (TAP)
boundary conditions
\begin{eqnarray}
  \label{bcTP}
  {\rm TP:}&&\quad\psi_1(-L/2)=\psi_2(L/2),\quad
\psi_2(-L/2)=\psi_1(L/2)\\
\label{bcTAP}
  {\rm TAP:}&&\quad\psi_1(-L/2)=-\psi_2(L/2),
\quad\psi_2(-L/2)=-\psi_1(L/2)
\end{eqnarray} amount to putting the system on a circle without
introducing a point where a
boundary is present: 
The kink solution $\phi_K(x)=\phi_0\tanh \frac{mx}{2}$ is invariant 
under the simultaneous transformation $x=L/2\to x=-L/2$ and 
$\phi_K\rightarrow
-\phi_K$. Thus the points $x=\pm L/2$ may be identified. 
The action for the susy kink
\be {\cal L}= -\frac{1}{2} (\partial_\mu \phi)^2 - \frac{1}{2} U^2(\phi) -
\frac{1}{2} \bar{\psi} \gamma^\mu \partial_\mu \psi - \frac{1}{2}
U'(\phi) \bar{\psi} \psi
\label{Lagrangian}
\ee with $U(\phi)=U(-\phi)$ is invariant under the transformation
\begin{equation}
  \label{eq:paritytranfrob}
  \phi\rightarrow-\phi\ , 
\psi\rightarrow e^{i\alpha}\gamma^3\psi \ \ ,
\end{equation} 
which is compatible with the Majorana condition for $\alpha = 0$ or $\pi$
whereas for Dirac fermions an arbitrary phase would be allowed.
Here we use a Majorana
representation of the
Dirac matrices with 
$\gamma^0=-i\sigma_2$, $\gamma^1=\sigma_3$, $\gamma^3=\sigma_1$. In
these terms, the TP and
TAP boundary conditions in (\ref{bcTP}),  (\ref{bcTAP}) are simply $\psi\to\pm
\gamma^3\psi$, clearly satisfying   (\ref{eq:paritytranfrob}). As a
consequence  there is no
visible boundary (meaning no locally observable discontinuity or cusp) at
$x=\pm L/2$.  Note that it is not necessary in these considerations to
assume that the
center of the kink is located at the point $x=0$.  That will be helpful
later on in
defining the parity operation in a simple manner, but for any other
purpose the matching
point for the transition or jump conditions (\ref{eq:paritytranfrob})
may be chosen
arbitrarily, as befits a locally invisible boundary.

The TP and TAP boundary conditions arise naturally if one begins with a  
kink-antikink
system with periodic (P) boundary conditions, and looks at the  values
of $\psi_1$, $\psi_2$
between the kink and antikink. One finds then that for P conditions for
the kink-antikink
system,  the fermions satisfy either TP or TAP conditions. In this 
article we also consider
a natural extension of these ideas: we begin with antiperiodic (AP)
boundary conditions for
the kink-antikink system,  and find then that if the  fermionic modes
are written as plane
waves $e^{-i(\omega t-kx)}$ far away from the kink-antikink system, then
in between the kink
and antikink they satisfy imaginary twisted periodic and antiperiodic
(iTP and iTAP)
boundary conditions
\begin{eqnarray}
  \label{bciTP} {\rm iTP}:\quad && 
\psi_1(-L/2)=i\psi_2(L/2),\quad\psi_2(-L/2)=i\psi_1(L/2)\\
\label{bciTAP} {\rm iTAP}:\quad &&   
\psi_1(-L/2)=-i\psi_2(L/2),\quad\psi_2(-L/2)=-i\psi_1(L/2)\;,
\end{eqnarray} 
where $\psi_{1,2}$ now refer to the fermionic mode functions
as opposed to the complete field.
If one prefers to avoid working with complex boundary
conditions for the 
Majorana fermions, one may take the real and imaginary parts of the 
distorted plane waves,
but this then leads to the {\em nonlocal} boundary conditions
$(-\partial_x^2+m^2)^{1/2}\psi_1(-L/2)=\pm(\partial_x-m)\psi_2(L/2)$ and
similar conditions
for $\psi_2(-L/2)$.  In the following we consider only the algebraic
boundary conditions
(\ref{bciTP}) and (\ref{bciTAP}). For periodic boundary conditions on
the kink-antikink
system,  one finds only the real boundary conditions for a single kink
given in (\ref{bcTP})
and (\ref{bcTAP}), whether one uses complex or real mode functions.

In the trivial sector, P and AP boundary conditions are invisible
boundary conditions, and,
having introduced iTP/iTAP it seems only natural to also include iP and
iAP boundary
conditions
\begin{eqnarray}
  \label{bciP} {\rm iP:}&&\quad  \psi_1(-L/2)=i\psi_1(L/2),\quad
\psi_2(-L/2)=i\psi_2(L/2)\nonumber\\
  \label{bciAP} {\rm iAP:}&&\quad   \psi_1(-L/2)=-i\psi_1(L/2),\quad
\psi_2(-L/2)=-i\psi_2(L/2).
\end{eqnarray}

With imaginary boundary conditions, 
one finds a generalized Majorana identity, in which the
adjoint of the field for one of the two boundary conditions is equal to
the field for the other boundary condition, so that only if one averages
over both conditions is it meaningful to describe the fermions as
Majorana particles.    

In Ref.~\cite{glv}, it was found that for a single kink one has to
consider suitable
averages over subsets of the mentioned
boundary conditions to obtain the correct
susy kink mass,
because  for particular  individual cases one encounters localized
boundary energy.  This
localized energy  is due to boundary conditions which distort the field
at the boundary and
may be called visible boundary conditions.   In the kink sector,
the P/AP and iP/iAP
boundary conditions are visible, whereas in the trivial sector, the
twisted versions are
visible.

To cancel out localized boundary energy, one needs to average over the
results  of
a twisted and an untwisted boundary condition. In this paper, we shall
show that there is a
reason to average also over the two twisted boundary conditions, because
a single (real)
twisted boundary condition breaks parity $\cal P$ (as well as $\cal T$),
giving rise to
delocalized momentum proportional to the ultraviolet cutoff, which
cancels only in the
average. (In the case of imaginary boundary conditions, a similar
phenomenon arises with
iP/iAP in the trivial sector.) This was overlooked in Ref.~\cite{glv},
which had
assumed parity-invariance for the spectrum and incorrectly claimed the
appearance of
delocalized energy.

One might expect that one can find other boundary conditions in the kink
sector which
preserve parity. Indeed, the invisible boundary conditions iTP, and iTAP
have a $\cal P$ and
$\cal T$ invariant spectrum, but instead violate $\cal C$ (and thus
$\cal CPT$), so that
these mode functions do not allow one to build a local quantum field
theory with Majorana
fields. Because $\cal C$ selects different locally invisible boundary
condition from $\cal
P$ and $\cal T$, it follows that there is no choice which preserves all
three symmetries
simultaneously.  This obstruction occurs despite the fact that the
action as a
local expression in Bose and Fermi fields is invariant under all the
symmetries.  Hence,
one encounters here a  phenomenon which we call with some  hesitation a
discrete symmetry
anomaly, induced  by the kink. 
%%%%%%%%%%%%%%%
There is no local counterterm which can remove this anomaly. One can, of
course, choose as boundary conditions $\psi=0$ in which case there are no
problems with the discrete symmetries, but then one has localized boundary
energy, and our aim here is to study the discrete symmetries in the
presence of invisible boundary conditions, which means with
the kink put on a circle.
%%%%%%%%%%%%%%%%

The possibility that a nontrivial structure of  spacetime can lead to
anomalies in  discrete
symmetries has been studied before. For example, in Ref.~\cite{KM}  a
$\cal CPT$ anomaly was
claimed to arise by  compactification of some dimensions of (3+1) 
spacetime. 

In our example, both a nontrivial space-time and a nontrivial field
topology is present. In
Ref.~\cite{rvw}, it was found that in 2+1 dimensions there arise
chiral  fermions living
on a susy kink domain wall;  these fermions are massless in 2+1
dimensions (their energy is
equal to the momentum along the domain wall) and they correspond to
fermionic zero modes of
the susy kink  in 1+1 dimensions. In this case the spectrum is again 
parity-nonsymmetric
(the massless fermions on the domain wall move in one direction but not
in the other) but
now this is not due to boundary  conditions but rather due to the
presence of the kink, in
accordance with the general results of
Refs.~\cite{Callan:1985sa,Gibbons:2000hg}.  In
Ref.~\cite{Affleck:as}  the connection between instantons and the
breaking of supersymmetry
and the discrete symmetries $\cal C,P,T$  was considered.

Our paper is organized as follows. In Sect.~\ref{sec:II} we discuss how the
symmetries $\cal C$, $\cal
P$ and 
$\cal T$ act on the  boundary conditions in the kink and in the trivial
sector. In Sect.~\ref{sec:III},
we work out the fermionic  spectra for the 16 sets of boundary
conditions (8 sets in the
kink sector, and 8  sets in the trivial sector). We also determine how
the total mass and
momentum of the kink depend on the choice of boundary conditions. We
regulate by mode 
regularization, i.e., requiring equal numbers of modes in the trivial
and kink sector,
counting  fermionic zero modes according to the rules derived in Ref.
\cite{glv}. In
Sect.~\ref{sec:IV} we comment on our results.

\section{Discrete symmetries and their implementation}
\label{sec:II}

For the single-particle Dirac Hamiltonian 
\begin{equation}
\label{H} H=i\hbar\sigma_1\partial_x +\hbar\sigma_2m\phi_K(x)/\phi_0 \
,\end{equation} one has
simple and unique representations of the three symmetry operations,
charge conjugation
${\cal C}$, parity ${\cal P}$, and  time reversal ${\cal T}$, which
leave this differential operator invariant.  ${\cal
C}$ at the
single-particle level is an antiunitary operation which  reverses the
sign of $H$, and
because in this representation $H$ is purely imaginary the transformation
is accomplished by
simple complex conjugation  of fermion wave functions:
${\cal C}={\cal K}$.    For ${\cal P}$, which must include the
transformation $x\to -x$, a
subtlety arises because this operation by itself turns the kink into  an
antikink. 
Therefore, in the kink sector,  one must require for the action of
parity on the classical
bosonic field $\phi_K(x)\to  -\phi_K(-x)=
  \phi_K(x)$.  In the kink background the combined transformation 
 reverses the derivative term but not the mass or Yukawa term in $H$,
and we 
 find for the action on fermion wave functions ${\cal P}=(x\to -x)\times
i\sigma_2$.  For the antiunitary ${\cal T}$ one needs an operation
including ${\cal K}$, but
it must leave $H$ invariant. To do this requires a matrix factor
anticommuting with $H$,
yielding ${\cal T}=\sigma_3{\cal K}$.
 Note that each of these discrete operations on fermion wavefunctions is
the  same in the
trivial sector as it is in the kink sector.  Of course, in the trivial
sector, the action of
parity on the (constant) classical background field
$\phi_0$ is simply to preserve it. Thus, to keep the background
invariant one  treats
 the background field as scalar in the trivial sector but pseudoscalar
in the kink sector.

While the discrete transformations can be defined consistently for the differential
operator, we still need to look at their effects on the matching or
boundary conditions. 
Let us write these conditions in a general form which covers all the
choices described above:
\begin{equation}\label{genbc}
\psi(x=-L/2)=\Gamma e^{i\alpha}\psi(x=+L/2) \ \ .
\label{alpha}
\end{equation}  The twisted boundary conditions which we now analyse
correspond to
$\Gamma=\gamma^3=\sigma_1$.  The conditions could be applied at any
point (see \cite{misha} for the details of the precise procedure), 
but let us choose symmetric  placement around the center of the
kink to make the
action of the parity symmetry as simple as possible.  Evidently we
obtain the four different
possibilities mentioned above by choosing $\alpha =0,\pi,\pi/2,-\pi/2$, 
respectively.  The
action of ${\cal C}$ takes $e^{i\alpha}$ to 
$(e^{i\alpha})^*$, so that only $\alpha=0,\pi$ (TP and TAP) are left unchanged.
 For parity, because of the interchange of left and right boundaries
along with the presence
of the matrix $\sigma_2$, one has $e^{i\alpha}\to -(e^{i\alpha})^{-1}$,
so that only
$\alpha=\pm \pi/2$ (iTP and iTAP) are left unchanged.
 For ${\cal T}$, the matrix $\sigma_3$ implies $e^{i\alpha}\to
-(e^{i\alpha})^*$,  and again
$\alpha=\pm \pi/2$ (iTP and iTAP) are left unchanged.\footnote{We use
the passive point of
view according to which we equate (\ref{genbc}) to
$\psi'(-L/2)=M\psi'(L/2)$ and solve for $M$.}

The purely real TP and TAP conditions commute with
${\cal C}$, but ${\cal T}$ and ${\cal P}$ each interchange TP with TAP.
Consequently, with
one of these conditions by itself only ${\cal C}$ holds:  It is possible
to choose wave
functions which are real, and a fermion field operator which is
Hermitean, but
(positive-energy) waves of positive and negative wavenumber
$k$ are not degenerate with each other.   This means that an implicit
assumption of
\cite{glv}, that the energy spectrum is the same for $k>0$ and $k<0$, is
 not correct
\cite{wim}.  In \cite{glv} the spectrum for negative $k$ was not
computed explicitly, and
this led to a false conclusion that the energy spectra for TP and TAP
are different.  In
fact, it is easy to check that for each solution with
$k$ of one sign for TP there is a degenerate solution with $k$ of the
opposite sign for
TAP.  A further assertion of \cite{glv} resulting from the assumed
difference in spectra is
that there exists a delocalized energy for either TP or TAP alone.  This
also is false
\cite{wim}, but
 as will be shown below there indeed is a delocalized quantity, a net 
 momentum proportional to the ultraviolet cutoff energy $\Lambda$.

On the other hand, with iTP and iTAP conditions, ${\cal P}$ and ${\cal
T}$ symmetries leave
the conditions invariant, but ${\cal C}$ interchanges them.  Once again,
to have all three
symmetries one must use an average over the two boundary conditions.
This time, if one just
chooses one of these boundary condition there is a difference in energy
spectrum from the
other boundary condition (but the spectra are each parity-symmetric). 
Now a new difficulty
arises, that it is impossible to write a Hermitean Majorana field,
because a positive energy
state with positive momentum does not have an equal negative energy
partner with negative
momentum.   A different way to reach the same conclusion is to consider
the operation ${\cal
CPT}$, which is a well-accepted symmetry for local quantum field
theory.\footnote{There has
been  recent interest in anomalous ${\cal CPT}$ violation in chiral
theories in 4 
dimensions \cite{KS,F} and in 2 dimensions \cite{KM}.  We consider the
present work (which
does not  include chiral gauge couplings) complementary to those
studies, but the chiral 
nature of the twisted boundary conditions suggests that there may be a 
connection to the
anomaly in explicitly chiral theories.} 

Evidently this symmetry leaves the field Hamiltonian density invariant
only for the TP and
TAP conditions, which therefore are the ones uniquely allowed as
consistent conditions in
quantum field theory.  For these conditions to achieve vanishing
delocalized momentum one
must average over TP and TAP, while for iTP and iTAP implementing ${\cal
CPT}$ symmetry
forces averaging over the two sets.  Thus the notion of averaging over
sets of boundary
conditions, as
 introduced in \cite{glv}, does have merit, but detailed claims
  in the original rationale for this construction needed major revision,
as we have just
described.

For completeness, we should examine the effects of the discrete
symmetries in the trivial
sector. Now, invisible boundary conditions have the unit matrix in place
of the matrix
$\sigma_1$.  One sees immediately that the P and AP conditions satisfy
all three discrete
symmetries, while iP and iAP do not satisfy any. This means that one
could implement the
discrete symmetries with either P or AP, but implementation for
imaginary conditions would
require both iP and  iAP.

To describe the discrete symmetries as transformations on the Majorana
field we need a
dictionary relating these transformations to those already discussed 
for  the  the single
particle wave functions.  For charge conjugation this is
\begin{equation} U_C\psi(x,t) U_C^{-1}=\psi^{\dagger}(x,t) \ \ ,
\end{equation} so that the Majorana condition becomes simply the
hermiticity or self 
adjointness of the field $\psi$.  Note that what had been an antiunitary
operation taking 
$H$  into its negative for the single-particle description now is a
unitary operation 
leaving the Hamiltonian density ${\cal H}(x.t)$ invariant.  This result depends
 critically on the  fact that ${\cal H}$ includes a commutator of $\psi$ with
$\psi^\dagger$, which reverses sign under charge conjugation.
For parity we have\footnote{As is the case for Majorana fermions
in 4 dimensions \cite{RYT}, ${\cal P}^2 = -1$.}
\begin{equation} U_P\psi(x,t) U_P^{-1}=i\sigma_2\psi(-x,t)  \ \ ,
\end{equation} identical with the single-particle rule. For time
reversal one finds the
greatest subtlety, because this operation remains antiunitary:
\begin{equation} V_T\psi(x,t) V_T^{-1} =\sigma_3\psi^*(x,-t) \ \ .
\end{equation} The subtlety has to do with defining complex conjugation
for the raising and
lowering operators $a^\dagger$ and $a$ appearing in the mode expansion
of the field.  The
simplest assumption is that this operation leaves the operators
invariant, but instead each
one could be multiplied by a different phase factor. In that case, the
phase factor would
have to be explicitly compensated in the  action of $V_T$ on each
raising or lowering
operator. It is easy to verify that these new definitions  are
consistent with the earlier
analysis of the relation between discrete symmetries and boundary
conditions, with the 
obvious proviso that the boundary conditions now are applied to the
field exactly as  they 
previously were applied to the wave functions.

The issues discussed here all arise because we are dealing with Majorana
 fermions. How
would the discussion change if one considered instead an $N=2$ theory, 
with Dirac
fermions?  Now the field $\psi$ no longer need be equivalent to its 
charge conjugate, so it
might seem that one could choose just one boundary condition  instead of
averaging over a
pair.  It is enticing to imagine that the Dirac  fermion  charge could
be coupled to a U(1)
gauge field, so that the phase $\alpha$ in  (\ref{alpha}) would reflect
a magnetic flux
threading the circle.  However, for no choice of $\alpha$ would the
spectrum obey all three
discrete symmetries, just as we found already; that deduction holds
regardless of the
assumption
$N=1$ or $N=2$.  Thus we still require a pair of boundary conditions if
the  symmetries all
are to be obeyed simultaneously.  In the $N=2$ theory however,
continuous  values of
$\alpha$ are allowed, and except for the values considered before any
other  would break all
three symmetries, as one would expect for arbitrary irrational  flux 
through the circle. 
The $N=2$ theory exhibits the Jackiw-Rebbi half-fermion  charge 
localized at the kink
\cite{JR}, and it is amusing that this is consistent with the
possibility of tunneling
between kink and antikink
\cite{Binosi:2000wy}, as the latter also  would possess charge one-half.
 The physical
interpretation of this analysis, when combined with what we saw earlier,
seems to be that
the problem of the kink on a circle `knows' that it really is half of
the kink-antikink
problem on a  doubled circle.  Thus the discrete symmetries which are
obeyed for half an
Aharonov-Bohm quantum of flux through the large circle also are obeyed
for one-quarter flux
through the small circle, but only when one averages over a suitable
pair (iTP and iTAP) of
boundary conditions.

\section{Mode number regularization of fermionic contributions
to the one-loop susy kink mass}
\label{sec:III}

We now turn to the explicit calculation of the fermionic contributions
to the susy kink mass at one-loop order in mode number regularization,
extending and partially correcting the results presented in Ref.~\cite{glv}.

%The classical Lagrangian of minimally $N=1$ susy kink models in
%1+1 dimensions is given by
%\be {\cal L}= -\frac{1}{2} (\partial_\mu \phi)^2 - \frac{1}{2} U^2(\phi) -
%\frac{1}{2} \bar{\psi} \gamma^\mu \partial_\mu \psi - \frac{1}{2}
%U'(\phi) \bar{\psi} \psi
%\label{Lagrangian}
%\ee
%where $\psi$ is a Majorana spinor, $\bar\psi=\psi^{\mathrm T} C$ and
%we use a Majorana representation of the Dirac matrices with 
%$\gamma^0=-i\sigma_2$, $\gamma^1=\sigma_3$, and $C=\sigma_2$.

The $\phi^4$-kink model corresponds to using
$U(\phi)=\sqrt{\lambda/2}(\phi^2-\phi_0^2)$
in the Lagrangian (\ref{Lagrangian}), but the following discussion
applies (mutatis mutandis) to other models such as sine-Gordon, 
where $U\propto\sin(\gamma\phi/2)$.

In the trivial vacuum, one has $U(\phi_0)=0$ and $U'(\phi_0)=
\sqrt{2\lambda}\phi_0=m$,
whereas with the nontrivial kink background field $\phi_K(x)=
\phi_0\tanh(m(x-x_0)/2)$
one has the Bogomol'nyi equation $U(\phi_K)=-\partial_x \phi_K$
and $U'(\phi_K)=m\phi_K/\phi_0$, leading to a fluctuation
equation for the fermionic mode functions governed by the
differential operator (\ref{H}).

The fermionic mode functions will be written
\be
\psi(x,t)=\left( \begin{array}{c}
\psi_1(x) \\
\psi_2(x) \end{array} \right) e^{-i \omega t}
\ee
so that the Dirac equation becomes
\be\label{Diraceq}
-i\omega \psi_1=(\partial_x-U')\psi_2,\quad
-i\omega \psi_2=(\partial_x+U')\psi_1.
\ee

The fermionic contribution to the one-loop quantum mass of a kink
is given by sums over zero-point energies according to
\be\label{Mfsum}
M_f^{(1)}=-{\hbar\over2}\left[ \sum \omega_K - \sum \omega_V \right]
+\Delta M_f
\ee
where the indices $K$ and $V$ refer to kink and trivial vacuum, respectively,
and $\Delta M_f$ is the fermionic contribution to the
counter-term due to renormalizing
the theory in the trivial vacuum. A minimal renormalization scheme
that can be chosen is to require that tadpoles vanish
and all other renormalization constants are trivial.\footnote{For
a thorough discussion of more general renormalization schemes in this
context see Ref.~\cite{rvw}.}
This gives \cite{glv}
\be
\Delta M_f=-{2\over3} \Delta M_b = -{m\hbar\over2\pi}
\int_{-\Lambda}^\Lambda {dk\over\sqrt{k^2+m^2}}.
\ee

In (global)\footnote{See Refs.~\cite{lmr,wim} for a local variant
which avoids the subtleties discussed here as well as allowing one
to calculate the local energy distribution.}  
mode regularization the spectrum of fluctuations about a kink (and
in the trivial vacuum) is discretized by considering an interval
of (large) length $L$ and choosing boundary conditions.
The sums in (\ref{Mfsum}) are then cut off
at a given large value $N$ of the number of modes, which 
according to the principle of mode regularization is chosen to be
the same in the trivial and in the kink sector.\footnote{The proper
regularization of these sums is a highly delicate matter. In particular,
a simple energy cutoff, which has frequently been employed
in the early literature \cite{Kaul:1983yt,Imbimbo:1984nq,Chatterjee:1984xh},
turns out to lead to results inconsistent with the exact integrability
of sine-Gordon models \cite{reb}. If, however, one uses a smooth
energy cut-off, one obtains an extra term in the mode sums
which is independent of the details of the smoothing, and this
then yields the correct result \cite{Litvintsev:2000is}.}

As argued in Ref.~\cite{glv}, this
requires {\em fixed} boundary conditions, meaning that they are
identical for the trivial and the kink sector.
But because invisible boundary conditions in one sector are visible ones
in the other, it becomes necessary to average over boundary conditions
such that boundary energies cancel in the average. 

The correct answer this average has to give is, as has been
established by a variety of methods 
\cite{schonfeld,Boya:1988zh,misha,Graham:1998qq,min,Litvintsev:2000is,glv,lmr,Bordag:2002dg,rvw},
\be\label{Mf1}
M_f^{(1)}=-M_b^{(1)}-{\hbar m\over 2\pi}
\ee
where $M_b^{(1)}$ is the bosonic contribution,
so that there is in total a nonvanishing negative correction
for the susy kink mass $M^{(1)}=M_f^{(1)}+M_b^{(1)}$
which is in fact entirely due to an interesting anomalous contribution
to the central charge operator \cite{min,Losev:2000mm,rvwprep}.

\subsection{Quantization conditions}

To explicitly compute the difference of the sums in Eq.~(\ref{Mfsum})
for the various boundary conditions discussed in Sect. 1, we have
to derive the quantization conditions on an interval of length $L$.
For ease of comparison with Ref.~\cite{glv}, Sect. VB, we let the spatial
coordinate run from $0$ to $L$ and put the center of the kink
at $x=L/2$.

We shall have to consider carefully both the discrete and
continuous\footnote{More precisely the part of the discretized spectrum that
becomes continuous in the limit $mL\to\infty$.} spectrum.

\subsubsection{Trivial sector}

If one puts
\be\label{psi1tr}
\psi_1 = e^{ikx}+ a e^{-ikx} \ee
then it follows from the Dirac equation (\ref{Diraceq}) with $U'\equiv m$ that
\be\label{psi2tr}
\psi_2=-[e^{i(kx+\frac{\theta}{2})} - a e^{-i(kx+\frac{\theta}{2})}]
\ee
where we define
$\theta $ such that 
\be\label{theta}
e^{i\frac{\theta}{2}}=\frac{k-im}{\omega},\quad \omega=\pm\sqrt{k^2+m^2}.
\ee
So $\theta=-2\arctan(m/k)$, but the branch of the arctan is fixed
such that (for positive frequencies $\omega$)
$\theta$ goes from $-2\pi$ to $0$ as $k$ runs from $-\infty$
to $+\infty$. This conforms with the definition
adopted in \cite{misha} but deviates from Ref.~\cite{glv}.
The definition (\ref{theta}) has the advantage of avoiding
explicit sign functions ${\rm sgn}(k)$ in the quantization conditions.
 
The quantization conditions for untwisted P and AP boundary conditions are
simply $kL=2\pi n$ and $kL=2\pi n+\pi$; iP and iAP
have $kL=2\pi n-\pi/2$ and $kL=2\pi n+\pi/2$, respectively. Notice
that iP and iAP in the trivial sector each have a
set of solutions which is not symmetrical under $k\to-k$.

The  
twisted boundary conditions read
$\psi_1(0)=\rho\psi_2(L)$ and 
$\psi_2(0)=\rho\psi_1(L)$, where $\rho=e^{i\alpha}=(+1,-1,+i,-i)$
for TP, TAP, iTP, and iTAP, respectively. 
Plugging these conditions into (\ref{psi1tr}) and (\ref{psi2tr}) 
and solving for $a$
gives 
\be\label{a}
\frac{-\rho e^{i(kL+\frac{\theta}{2})}-1}{-\rho e^{-i(kL+\frac{\theta}{2})}+1}
=a=\frac{-e^{i
\frac{\theta}{2}}- \rho e^{ikL}}{\rho e^{-ikL}-e^{-i\frac{\theta}{2} }}\;.
\ee
Multiplying out, this gives
\be
(\rho^2-1)(e^{i\frac{\theta}{2}}+e^{-i\frac{\theta}{2}})
=2\rho(e^{ikL}-e^{-ikL}).
\ee

For $\rho^2=1$, (TP and TAP),
this is equivalent to $\sin kL = 0$, i.e. $kL=\pi n$,
with $n\ne 0$, because $n=0$ corresponds to the trivial solution 
$\psi_1=\psi_2=0$.

Imaginary twisted periodic/antiperiodic boundary conditions (iTP/iTAP) have
$\rho^2=-1$, and one finds for $\rho=\pm i$ the two sets of solutions
$\mathrm a)\; kL=2\pi n-\frac{\theta}{2}\pm \frac{\pi}{2}$,\
$\mathrm b)\; kL=2\pi n+\frac{\theta}{2}\pm \frac{\pi}{2}$.
(For these conditions the numerator and denominator on one side
of (\ref{a}) vanish, but not on the other side.)
To every solution with $k$ there is one with $-k$, but the two
correspond to the same solution (up to normalization) so it suffices
to consider $k\ge0$;
$k=0$ has again $a=-1$ such that $\psi_1=\psi_2=0$ everywhere and therefore
must not be counted.

There are also potentially {\em zero modes}, $\omega=0$, and
{\em approximately-zero modes}, $\omega\approx0$, which have
to be treated separately. For $\omega=0$, the solutions to the
Dirac equation read
\be
\label{zeromodestr}
\psi_I = \left(
\begin{array}{c}
 {a_1}{e^{-mx}} \\  
 a_2 e^{mx} 
\end{array}
\right),
\ee
where $a_1$ and $a_2$ are determined by the boundary conditions.

Only TP and TAP give nontrivial solutions for $a_1$ and $a_2$ and
thus are compatible with these solutions. There is
one such zero mode for each of these boundary conditions.

The imaginary twisted boundary conditions
iTP/iTAP on the other hand have {\em almost-zero} modes
with
energy $\omega^2 \to 4m^2 e^{-2mL}$ for $mL\to\infty$,
with the positive-frequency solution satisfying iTP,
and the negative-frequency one satisfying iTAP.
To verify this, one can use the ansatz
\be
\psi_1=e^{-\kappa x}+a e^{\kappa x},\quad
-i\omega\psi_2=(m-\kappa)e^{-\kappa x}+a (m+\kappa) e^{\kappa x}
\ee
with $\omega^2=m^2-\kappa^2$ and make the approximation
$\kappa\approx m$ which becomes valid in the limit $mL\to\infty$.

The untwisted boundary conditions P, AP, iP, and iAP
have neither zero nor almost-zero modes in the trivial sector.

\subsubsection{Kink sector}

In the kink sector, one has asymptotic expressions
\be
\psi_1= \left\{ 
\begin{array}{c} e^{i(kx-\frac{\delta}{2})} + a e^{-i(kx-\frac{\delta}{2})},
\hspace{1cm}x \approx 0 \\ e^{i(kx+\frac{\delta}{2})} + a
e^{-i(kx+\frac{\delta}{2})}, \hspace{1cm}x \approx L \\
\end{array}
\right.
\ee
\be
\psi_2= -\left\{ 
\begin{array}{c} e^{i(kx-\frac{\delta}{2}-\frac{\theta}{2})} - a
e^{-i(kx-\frac{\delta}{2}-\frac{\theta}{2})},
\hspace{1cm}x \approx 0 \\ e^{i(kx+\frac{\delta}{2}+\frac{\theta}{2})} - a
e^{-i(kx+\frac{\delta}{2}+\frac{\theta}{2})},
\hspace{1cm}x \approx L \\
\end{array}
\right.
\ee
where $\delta=-2\arctan(3mk/(m^2-k^2))$ is the phase shift function
also appearing for bosonic fluctuations. So $\psi_1$ behaves as the latter,
while $\psi_2$ has a modified phase shift $\delta+\theta$.

For $\delta(k)$ we adopt the convention that $\delta(k\to\pm\infty)\to0$
so that there is a discontinuity at $k=0$ which in accordance with
Levinson's theorem is $2\pi$ times the number of bound states.
For $\theta$ we however keep the definition of Eq.~(\ref{theta}), which
has the advantage of avoiding a separate treatment of positive and negative
values of $k$.

\begin{figure}
\includegraphics[bb=175 235 460 485,width=9cm]{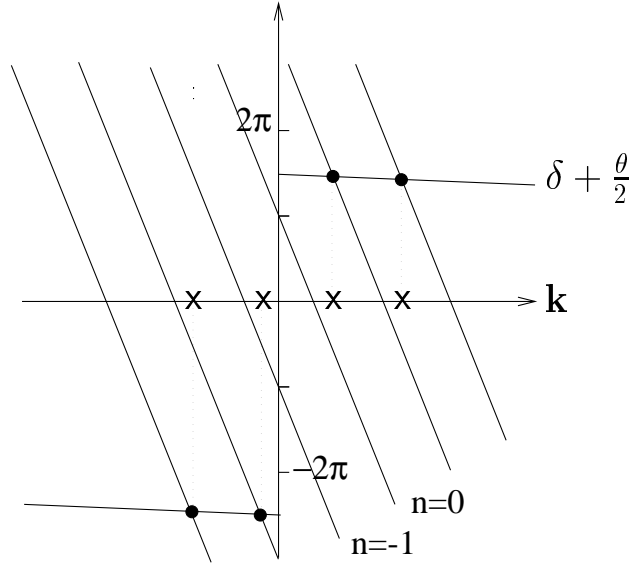}
\caption{\label{figTP}The quantization conditions for the fermionic modes
in the case of TP boundary conditions obtained from
solving $\delta+\frac{\theta}{2}=2\pi n+\pi-kL$
for positive $\omega$. 
The spectrum is clearly not invariant under $k\to-k$.
}
\end{figure}

We begin with discussing the untwisted boundary conditions.

The (real) P and AP conditions can be satisfied either
for a) $a=1$ and $kL=2\pi n+\pi-\delta-\theta$ or
b) $a=-1$ and $kL=2\pi n+\pi-\delta$, where only positive $n$ need
to be considered to obtain a complete set of solutions and
solutions with $k=0$ have to be excluded, for they correspond to
$\psi_1=\psi_2=0$.
Because these quantization conditions involve only
$e^{i\theta}$ rather than $e^{i\theta/2}$,
in this (and only in this) case it would make no difference to
define $\theta$ such as to vanish for
$k\to\pm\infty$, as done for example in Ref.~\cite{reb} (which
obtained an incorrect result for the susy kink mass only because
there is a localized boundary energy contribution \cite{glv}, as
we shall see shortly). 

The imaginary untwisted boundary conditions iP and iAP on the other hand
have identical quantization conditions, which are given
by the two sets
a) $kL=2\pi n+{\pi\over2}-\delta
-{\theta\over2}$, $n\ge1$, b) $kL=2\pi n-{\pi\over2}-\delta
-{\theta\over2}$, $n\ge2$. Again, only positive $n$ need
to be considered since (in contrast to iP/iAP in the trivial sector)
$k\to-k$ does not lead to further independent solutions.

Turning now to the twisted boundary conditions, the
TP ones lead to $kL=2\pi n+\pi-\delta-\theta/2$. As shown
in Fig.~\ref{figTP}, this has solutions for all $n$ except
$n=0, -1$, and the set of these solutions is not symmetric
under $k\to-k$. The solutions generated by the latter
transformation instead obey TAP boundary conditions, which
require $kL=2\pi n-\delta-\theta/2$.

The imaginary twisted boundary conditions iTP/iTAP differ from
TP/TAP simply by an additional term $-\pi/2$ on the r.h.s.
of the quantization conditions (for positive-frequency
solutions). For iTP the exemptions
are $n=0, -1$ as with TP. For iTAP, $n=0$ has to be excluded,
while $n=\pm 1$ corresponds to
the threshold mode $k=0$,
$\omega=m$, which is proportional to
$(\psi_1,\psi_2)=(1-3\tanh^2(mx/2),-2i\tanh(mx/2))$, and thus consistent
with iTAP boundary conditions (it does not appear in any of the
other boundary conditions). Thus $n=\pm1$ has to be counted only once.

In contrast to TP/TAP, the sets of allowed $k$-values for iTP and iTAP 
are each symmetric
under $k\to-k$ (while the corresponding solutions are linearly independent),
%in accordance with the discussion of Sect.~\ref{sec:II}.
but a positive-frequency solution with
momentum $k$ for iTP or iTAP has a negative-frequency partner
only for the other of the two imaginary twisted boundary conditions.

\begin{table}
\caption{\label{tab} Summary of fermionic quantization conditions, numbered
in conformity with Ref.~\cite{glv} where applicable, and
the number of (almost-)zero modes ($n_z$) in each case.
An upper index $\pm$ to the number $n_z$ indicates that
these modes are only almost-zero modes; an index $+$ or $-$
indicates that only the positive or negative frequency mode, respectively,
is compatible with the given boundary condition (b.c.).}
\begin{ruledtabular}
\begin{tabular}{lllll}
$i)$ & b.c. & sector & $k_{i)}L$ & $n_z$ \\
\hline
1) & P & trivial & $2\pi n$, all $n$ & 0 \\
2) & AP & trivial & $2\pi n+\pi$, all $n$ & 0 \\
3) & P & kink & a) $2\pi n-\delta-\theta$, $n\ge 1$ & 2 \\
 & & & b) $2\pi n-\delta$, $n\ge 2$ &  \\
4) & AP & kink & a) $2\pi n+\pi-\delta-\theta$, $n\ge 1$ & $2^\pm$ \\
 & & & b) $2\pi n+\pi-\delta$, $n\ge 1$ &  \\
\hline
1') & iP & trivial & $2\pi n-\pi/2$, all $n$ & 0 \\
2') & iAP & trivial & $2\pi n+\pi/2$, all $n$ & 0 \\
3')=4') & iP/iAP & kink & a) $2\pi n+\pi/2-\delta-\theta/2$, $n\ge1$ & 
$2^\pm$ \\
 & & & b) $2\pi n+\pi/2-\delta-\theta/2$, $n\ge 2$ &  \\
\hline
5)=6) & TP/TAP & trivial & a) $2\pi n$, $n\ge 1$ & 1 \\
 & & & b) $2\pi n+\pi$, $n\ge 0$ &  \\
7) & TP & kink & $2\pi n+\pi-\delta-\theta/2$, all $n$, $n\not=0,-1$ & 1 \\
8) & TAP & kink & $2\pi n-\delta-\theta/2$, all $n$, $n\not=0,-1$ & 1 \\
\hline
5') & iTP & trivial & a) $2\pi n+\pi/2-\theta/2$, $n\ge 0$ & $1^+$ \\
 & & & b) $2\pi n+\pi/2+\theta/2$, $n\ge 1$ &  \\
6') & iTAP & trivial & a) $2\pi n-\pi/2-\theta/2$, $n\ge 1$ & $1^-$ \\
 & & & b) $2\pi n-\pi/2+\theta/2$, $n\ge 1$ &  \\
7') & iTP & kink & $2\pi n+\pi/2-\delta-\theta/2$, all $n$, $n\not=0,-1$ 
& $1^+$ \\
8') & iTAP & kink & $2\pi n-\pi/2-\delta-\theta/2$, all $n$, $n\not=0,+1$ 
& $1^-$ \\
\end{tabular}
\end{ruledtabular}
\end{table}

For the counting of modes in the next subsection we also need to know how
many zero modes there are for each boundary condition in the
kink sector. For
real boundary conditions these have been discussed in Ref.~\cite{glv}
and are recapitulated in Table \ref{tab}, which summarizes
the results of this subsection. The imaginary
boundary conditions iP and iAP each have a pair of approximately-zero
modes; however, for iTP there is only one approximately-zero mode
with positive frequency, while the complex conjugated negative-frequency
mode satisfies iTAP boundary conditions.
(For iTP and iTAP boundary conditions, one can take $\psi_1$ real
and $\psi_2$ purely imaginary as this is consistent with the Dirac
equation, while for iP and iAP both $\psi_1$ and $\psi_2$
are complex combinations of two real solutions.)

%iTP and iTAP share a pair of approximately-zero
%modes with the positive-frequency solution obeying iTP
%and the negative-frequency one obeying iTAP.

Finally, in the kink sector there is one bound state
with energy squared $\omega_B^2={3\over4}m^2$. One
can verify that on a finite interval it is possible
to satisfy any of the boundary conditions considered
by slightly increasing or decreasing the value of $\kappa_B$ in
$\omega^2_B=m^2-\kappa^2_B$.
This is easy to see for P, AP, iTP, and iTAP boundary
conditions where the mode functions $\psi_1$ and $\psi_2$
are antisymmetric and symmetric around the kink center,
respectively; for TP, TAP, iP, and iAP, we have
verified the compatibility of the boundary conditions
numerically. By contrast, the situation is more
complicated for the zero modes, because there
$\kappa_0$ can only be decreased from its maximal value
$\kappa_0=m$. Increasing $\kappa_0$ would turn $\omega^2$ negative,
but the Hamiltonian (\ref{H}) is self-adjoint with a Hermitean inner
product.

\subsection{Mode sums}

\subsubsection{Real boundary conditions}

Evaluating (\ref{Mfsum}) with an
equal number of modes in the trivial and in the kink sector, one
thus obtains for P and AP boundary conditions \cite{Uchiyama:1986gf,reb,glv}
\bea
M^{(1)}_f({\rm P})&=&  \frac{\hbar}{2} \sum_{n=-N}^{N} \omega_{1)} -
 \frac{\hbar}{2} \sum_{n=1}^{N} \omega_{3\mathrm a)} - \frac{\hbar}{2} 
\sum_{n=2}^{N} \omega_{3\mathrm b)} - 0 - \frac{\hbar \omega_B}{2} + \Delta M_{f}\nn\\ 
&=&- \frac{\hbar \omega_B}{2}+{\hbar m } + \hbar \int_0^{\Lambda}
\frac{dk}{2 \pi}
\omega^\prime \left( \delta + \frac{\theta}{2} \right) + \Delta M_{f}
\eea
and
\be
M^{(1)}_f({\rm AP})= {\hbar} \sum_{n=0}^{N} \omega_{2)} -
 \frac{\hbar}{2} \sum_{n=1}^{N} \omega_{4\mathrm a)} - \frac{\hbar}{2} \sum_{n=1}^{N}
\omega_{4\mathrm b)} - 0 -
\frac{\hbar \omega_B}{2} + 
\Delta M_{f}=M^{(1)}_f({\rm P}),
\ee
where the sums for the trivial sectors are written first, with
$\omega_{i)}=\sqrt{k_{i)}^2+m^2}$ according to Table \ref{tab};
explicit zeros indicate the presence of (almost-)zero modes.
This leads to
\be
M^{(1)}_f({\rm P}) = M^{(1)}_f({\rm AP})= M_f^{(1)}+{\hbar m\over 4}\;,
\ee
implying that there is a finite amount of boundary energy equivalent
to the contribution of one half of that of a low-lying
continuum mode. Since P and AP are invisible boundary conditions
in the trivial sector, this must be attributed to the kink sector.

For fixed TP and TAP boundary conditions, we find (correcting Ref.~\cite{glv})
\bea
M^{(1)}_f({\rm TP})&=& \frac{\hbar}{2} \sum_{n=1}^{N} \omega_{5\mathrm a)} +
 \frac{\hbar}{2} \sum_{n=0}^{N} \omega_{5\mathrm b)} -  \frac{\hbar}{2}\sum_{n=1}^{N}
\omega_{7)} -  \frac{\hbar}{2}\sum_{n=-2}^{-(N+1)}
\omega_{7)}  - \frac{\hbar
\omega_B}{2} + \Delta M_{f} \nn\\
&=&  - \frac{\hbar \omega_B}{2}+\frac{\hbar m }{2} + \hbar
\int_0^{\Lambda} \frac{dk}{2 \pi}
\omega^\prime \left( \delta + \frac{\theta}{2} \right) + \Delta M_{f}
= M_f^{(1)}-{\hbar m\over 4}
\eea
and
\bea
M^{(1)}_f({\rm TAP})&=& \frac{\hbar}{2} \sum_{n=1}^{N} \omega_{6\mathrm a)} +
 \frac{\hbar}{2} \sum_{n=0}^{N} \omega_{6\mathrm b)} -  \frac{\hbar}{2}\sum_{n=1}^{N}
\omega_{8)} -  \frac{\hbar}{2}\sum_{n=-2}^{-(N+1)}
\omega_{8)}  - \frac{\hbar
\omega_B}{2} + \Delta M_{f}
\nn\\
&=&  - \frac{\hbar \omega_B}{2}+\frac{\hbar m }{2} + \hbar
\int_0^{\Lambda} \frac{dk}{2 \pi}
\omega^\prime \left( \delta + \frac{\theta}{2} \right) + \Delta M_{f}
= M^{(1)}_f({\rm TP}).
\eea

TP/TAP are invisible boundary conditions
in the kink sector, so that any boundary energy must now be
attributed to the trivial sector. As one can see, it has
equal magnitude but opposite sign than in the results for P/AP, 
in agreement with the
discussion in Ref.~\cite{glv}. (Twisting the fermions from
P in the trivial sector to TP in the kink sector, the
localized boundary energy does not change.) However, because
$M^{(1)}_f({\rm TAP})=M^{(1)}_f({\rm TP})$, there is no
delocalized boundary energy in the sense of Ref.~\cite{glv}.

Taking the average of the results of one of the untwisted and one of
the twisted boundary conditions eliminates the localized boundary
energy and yields the correct result (\ref{Mf1}). 

In Ref.~\cite{misha} it was found that mode number
regularization with the completely invisible ``topological''
boundary conditions of P in the trivial sector
and TP in the kink sector produces the correct finite part, but
leaves an infinite (but $m$-independent) term corresponding to
the contribution of one half of that of a continuum mode with $k=\Lambda$.
The latter is removed by the derivative regularization method
proposed in Ref.~\cite{misha}. For mode regularization to give
finite results it is
crucial to have fixed boundary conditions. The localized boundary energies
that this produces has then to be eliminated by averaging over
one twisted and one untwisted boundary condition.

However, taking either TP or TAP for the twisted boundary condition, 
parity $\cal P$ is not a symmetry
and thus the one-loop correction to the momentum in the kink
sector need not be zero.

The momentum operator is diagonal asymptotically far away from the
kink, and one obtains for TP
\bea
P_f^{(1)}({\rm TP})&=&
{\hbar\over 2L} \left(\sum_{n=1}^{N}+\sum_{n=-2}^{-(N+1)}\right)
[2\pi n+\pi-\delta-\theta/2]\nn\\
&=&{\hbar\over 2}\int_0^\Lambda {dk\over 2\pi}
[-\delta-\theta/2]+
{\hbar\over 2}\int_{-\Lambda}^0{dk\over 2\pi}[-\delta-\theta/2]
=+{\hbar\over 4}\Lambda
\eea
and, for TAP,
\bea
P_f^{(1)}({\rm TAP})&=&
{\hbar\over 2L} \left(\sum_{n=1}^{N}+\sum_{n=-2}^{-(N+1)}\right)
[2\pi n-\delta-\theta/2]\nn\\
&=&{\hbar\over 2}\int_0^\Lambda {dk\over 2\pi}
[-\delta-\theta/2]+
{\hbar\over 2}\int_{-\Lambda}^0{dk\over 2\pi}[-2\pi-\delta-\theta/2]
=-{\hbar\over 4}\Lambda\;.
\eea
Both results correspond to the contribution of one-half of a
high-energy mode $|k|=\Lambda$, but with opposite sign. So there is
an infinite amount of ``delocalized momentum'', which cancels only
in the average over TP and TAP.

\subsubsection{Imaginary boundary conditions}

As discussed in Sect.~\ref{sec:II}, the imaginary versions of the
above boundary conditions have the problem that each of iP, iAP,
iTP, and iTAP separately break $\cal C$ and make it impossible
to define Majorana quantum fields. In fact, $\cal CPT$ is equally
violated.

Nevertheless, it may make sense to consider these boundary conditions
in an averaged sense. Summing over positive frequencies only one has
for iP
\bea
M^{(1)}_f({\rm iP})&=&\frac{\hbar}{2} \sum_{n=-N}^{N} \omega_{1')}-  
\frac{\hbar}{2}\sum_{n=1}^{N}
\omega_{3\mathrm a')} - \frac{\hbar}{2}\sum_{n=2}^{N}
\omega_{3\mathrm b')} - 0 - \frac{\hbar
\omega_B}{2} + \Delta M_{f}
\nn\\&=&  - \frac{\hbar \omega_B}{2}+2\times \frac{\hbar m }{2} 
+ {\hbar}
\int_0^{\Lambda} \frac{dk}{2 \pi}
\omega^\prime \left( \delta + \frac{\theta}{2} \right) + \Delta M_{f}
= M_f^{(1)}+{\hbar m\over 4},
\eea
and the same for $M^{(1)}_f({\rm iAP})$ because $\sum_{-N}^{N} \omega_{1')}=
\sum_{-N}^{N} \omega_{2')}$ and (3')=(4') according to Table \ref{tab}.
The iP/iAP results for the one-loop energies 
thus coincide with the corresponding results for P/AP.

Analogously, for iTP one obtains
\bea M^{(1)}_f({\rm iTP})&=& 0+\frac{\hbar}{2} \sum_{n=0}^{N} \omega_{5\mathrm a')} +
 \frac{\hbar}{2} \sum_{n=1}^{N} \omega_{5\mathrm b')} -  2\times
\frac{\hbar}{2}\sum_{n=1}^{N}
\omega_{7')} - 0 - \frac{\hbar
\omega_B}{2} + \Delta M_{f}
\nn\\&=&  - \frac{\hbar \omega_B}{2}+\frac{\hbar m }{2} 
%+\frac{\hbar}{2} \sqrt{\Lambda^2+m^2}
+ {\hbar}\int_0^{\Lambda} \frac{dk}{2 \pi}
\omega^\prime \left( \delta + \frac{\theta}{2} \right) + \Delta M_{f}
%$$
%$$
= M_f^{(1)}-{\hbar m\over 4} %+\frac{\hbar}{2} \sqrt{\Lambda^2+m^2},
\eea
and for iTAP
\bea M^{(1)}_f({\rm iTAP})&=& \frac{\hbar}{2} \sum_{n=1}^{N} \omega_{6\mathrm a')} +
 \frac{\hbar}{2} \sum_{n=1}^{N} \omega_{6\mathrm b')} -  
\frac{\hbar m }{2} - 2\times
\frac{\hbar}{2}\sum_{n=2}^{N}
\omega_{8')} - \frac{\hbar
\omega_B}{2} + \Delta M_{f}
\nn\\&=&  - \frac{\hbar \omega_B}{2}+\frac{\hbar m }{2} 
+ {\hbar}
\int_0^{\Lambda} \frac{dk}{2 \pi}
\omega^\prime \left( \delta + \frac{\theta}{2} \right) + \Delta M_{f}
= M^{(1)}_f({\rm iTP}).
\eea

Although $\cal C$ is broken, the two results coincide, so there is
still no delocalized boundary energy in the sense of Ref.~\cite{glv}.

Because $\cal P$ is intact with either iTP or iTAP, there is
also no delocalized momentum as with real twisted boundary conditions.
However, iP/iAP in the trivial sector now break $\cal P$ (whereas
the kink sector is symmetric under $k\to-k$), and one finds
that there is delocalized momentum associated with the trivial
sector,
\be
P_f^{(1)}({\rm iP})={\hbar\over 2L}\sum_{n=-N}^N (2\pi n-{\pi\over2})
=-{\hbar\over 4}\Lambda
\ee
and
\be
P_f^{(1)}({\rm iAP})={\hbar\over 2L}\sum_{n=-N}^N (2\pi n+{\pi\over2})
=+{\hbar\over 4}\Lambda,
\ee
which again corresponds to the contribution of one-half of a
high-energy mode $|k|=\Lambda$ for iP and iAP separately, but
with opposite sign.

Thus, averaging over the results of the mode sums for
all four imaginary boundary conditions
removes both localized boundary energies and delocalized momentum.
In fact,
only in such an average one effectively removes
also the obstruction to the Majorana condition (and $\cal CPT$)
that positive and negative frequency modes have different spectra.

Curiously enough, the necessity to consider
iTP and iTAP together in order to have at least
effectively no violation of $\cal C$ and $\cal CPT$
means that the threshold mode $k=0$, which only
appears under iTAP boundary conditions, is in the
average counted like half a mode. In Ref.~\cite{Graham:1998qq},
in a different regularization method, threshold modes
had to be treated explicitly as modes to be counted
only half.

\section{Discussion}
\label{sec:IV}

We have considered the susy kink on a circle by introducing
so-called invisible boundary conditions as proposed earlier
in Refs.~\cite{misha,glv}. We then analysed how the discrete
symmetries $\cal C$, $\cal P$, and $\cal T$ act on these
boundary conditions. We found that no single set of
locally invisible boundary conditions preserved all
three discrete symmetries. The real boundary conditions
TP and TAP preserved $\cal CPT$, but break both
$\cal P$ and $\cal T$. The imaginary variants iTP and iTAP
on the other hand respect $\cal P$ and $\cal T$, but
violate $\cal C$ and therefore even $\cal CPT$, so that
these boundary conditions cannot be used for local
quantum field theory, although this obstruction is
effectively removed by
averaging over iP and iAP, or iTP and iTAP.
The cancellation of local boundary energy in the 
mode regularization scheme requires averaging over
the results obtained with
one twisted and one untwisted boundary condition,
where these conditions have to be used both in the
trivial and in the kink sector.

For compatibility with the Euler-Lagrange variational
principle, one should require that boundary terms due
to partial integrations cancel. In our case these
``boundary field equations'' read
\bea
&&\psi_1(-L/2)\delta\psi_2(-L/2)+\psi_2(-L/2)\delta\psi_1(-L/2)\nn\\=&&
\psi_1(L/2)\delta\psi_2(L/2)+\psi_2(L/2)\delta\psi_1(L/2).
\eea
It is easy to see that the real boundary conditions
P, AP, TP, and TAP all satisfy this requirement, but
the imaginary versions iP, iAP, iTP, and iTAP
each violate it.

This means that none of the imaginary boundary conditions
can be used in a Lagrangian formulation with
Majorana fermions, although the Hamiltonian (\ref{H})
with a Hermitean inner product is still self-adjoint. 
The same conclusion was reached
by looking at the spectrum (derived from bulk field
equations and imposing the boundary conditions). The
problem with imaginary boundary conditions then turned
out to be that for a given momentum $k$ and positive
frequency $\omega$ there is no corresponding mode
in the spectrum with $-k$ and $-\omega$, and
no Majorana field can be built.

To avoid this problem, one would have to switch to
complex fermions by giving up supersymmetry, as
in the original Jackiw-Rebbi model \cite{JR}, or go
to $N$=2 susy models. Neither possibility has been explored
in this paper.

%%%%%%%%
We summarize our assertions about averaging over invisible boundary
conditions to restore all three discrete symmetries.  In the trivial
sector, one may average over P and AP or iP and iAP, or both sets. 
However, because P and AP separately obey all symmetries, there is no
need to average if one chooses one of these real periodic boundary
conditions.   In the kink sector, one may average over TP and TAP or iTP
and iTAP, or both sets.  Any of these is an acceptable method to restore
the symmetries, but this time there is no single boundary condition
which simultaneously satisfies all three, so that averaging over at
least one pair is necessary.  That fact is the main point of our work. 

The idea that one must average over a set of boundary conditions to
restore a symmetry is known in string theory, where the spinning string
maintains modular invariance (large general coordinate transformations)
and unitarity and supersymmetry only if one sums over all spin structures 
(the requirement that fermions on a closed surface are periodic or antiperiodic
in spacelike or timelike directions) \cite{GSW}.
%%%%%%%%%

We close with some speculative remarks.
The fact that no {\em locally invisible} boundary condition
for the fermionic quantum fluctuations satisfies
all three symmetries $\cal C$, $\cal P$, and $\cal T$
simultaneously, whereas the classical action is
$\cal C$, $\cal P$, and $\cal T$ invariant, suggests
that we are dealing with a discrete anomaly. The origin
of this effect is the global structure (analogous
to a M\"obius strip in our case \cite{misha}), whereas
the usual chiral anomaly is a local effect.
Clearly, one should not confuse this with the anomalies
due to instantons, where the effective action contains
terms of the form $\psi^4+\bar\psi^4$; these preserve
parity but break chiral invariance.

Whether or not the striking loss of
simultaneous $\cal C$, $\cal P$, and $\cal T$ invariance 
should be called an anomaly 
in the sense 
of the chiral anomaly, it certainly satisfies the
definition of an anomaly 
as a `clash of quantum consistency conditions' \cite{RJ}. 

\acknowledgments

This research was supported in part by the National Science Foundation,
Grant Nos. PHY00-98527, PHY99-07949, and by the
Austrian Science Foundation FWF, Project No. P15449.


\begin{thebibliography}{99}
\bibitem{RYT}
G. Racah, Nuovo Cim. {\bf 14}, 322 (1937);
C.N. Yang and J. Tiomno, Phys. Rev. {\bf 79}, 495 (1950).
%See, for example, B.J. Kayser and A.S. Goldhaber, Phys. Rev. D
% {\bf 28}, 2341 (1983).
\bibitem{BK}
For more recent work see, for example, M. Nowakowski, Phys. Rev.
D {\bf 64}, 116001 (2001)
and references therein.
\bibitem{misha}
H. Nastase, M. Stephanov, P. van Nieuwenhuizen and
A. Rebhan,
Nucl. Phys. {\bf  B542}, 471 (1999).
\bibitem{glv}
A.S. Goldhaber, A. Litvintsev and P. van
Nieuwenhuizen,
Phys. Rev. D {\bf 64}, 045013 (2001).
\bibitem{KM} F.R. Klinkhamer and J. Nishimura, Phys. Rev. D {\bf 63}, 097701
(2001);
F.R. Klinkhamer and C. Mayer, Nucl. Phys. {\bf B616}, 215 (2001).
\bibitem{rvw}
A. Rebhan, P. van Nieuwenhuizen and R. Wimmer, New J. Phys. {\bf 4},
31 (2002).
\bibitem{Callan:1985sa}
C.~G. Callan and J.~A. Harvey,
\newblock Nucl. Phys. {\bf B250}, 427 (1985).
\bibitem{Gibbons:2000hg}
G.~W. Gibbons and N.~D. Lambert,
\newblock Phys. Lett. B {\bf 488}, 90 (2000). %, arXiv:hep-th/0003197.
\bibitem{Affleck:as}
I.~Affleck, J.~A.~Harvey and E.~Witten,
%``Instantons And (Super)Symmetry Breaking In (2+1)-Dimensions,''
Nucl.\ Phys.\ {\bf B206}, 413 (1982).
%
\bibitem{wim}
R. Wimmer,
{\it Quantization of supersymmetric solitons}
 (Diploma thesis, Vienna, Tech. U.), %. TUW-01-22, Sep 2001. 115pp, 
hep-th/0109119. 
\bibitem{KS}  F.R. Klinkhamer and J. Schimmel, {\it CPT anomaly: a
rigorous 
result in four dimensions}, hep-th/0205038, and earlier works cited therein.
\bibitem{F}  K. Fujikawa, M. Ishibashi, and H. Suzuki, JHEP {\bf 0204},
046 
(2002). This paper explicitly discusses CP violation.
\bibitem{JR} 
R. Jackiw and C. Rebbi, Phys. Rev. D {\bf 13}, 3398 (1976).
\bibitem{Binosi:2000wy}
D.~Binosi, M.~A.~Shifman and T.~ter Veldhuis,
%``Leaving the BPS bound: Tunneling of classically saturated solitons,''
Phys.\ Rev.\ D {\bf 63}, 025006 (2001).
%[arXiv:hep-th/0006026].
%%CITATION = HEP-TH 0006026;%%
\bibitem{lmr}
A.S. Goldhaber, A. Litvintsev and P. van
Nieuwenhuizen,
{\it Local Casimir energy for solitons},
hep-th/0109110. 
\bibitem{Kaul:1983yt}
R.~K. Kaul and R.~Rajaraman,
\newblock Phys. Lett. B {\bf 131}, 357 (1983).
\bibitem{Imbimbo:1984nq}
C.~Imbimbo and S.~Mukhi,
\newblock Nucl. Phys. {\bf B247}, 471 (1984).

\bibitem{Chatterjee:1984xh}
A.~Chatterjee and P.~Majumdar,
\newblock Phys. Rev. D {\bf 30}, 844 (1984);
\newblock Phys. Lett. B {\bf 159}, 37 (1985).
\bibitem{reb}
A. Rebhan and P. van Nieuwenhuizen,
Nucl. Phys. {\bf  B508}, 449 (1997).
\bibitem{Litvintsev:2000is}
A.~Litvintsev and P.~van Nieuwenhuizen,
{\it Once more on the BPS bound for the SUSY kink},
hep-th/0010051.
%%CITATION = HEP-TH 0010051;%%
\bibitem{schonfeld}
J.F. Schonfeld,
Nucl. Phys.  {\bf B161}, 125 (1979).
\bibitem{min}
M. Shifman, A. Vainshtein and M. Voloshin, Phys. Rev. D {\bf 59}, 045016
(1999).
\bibitem{Boya:1988zh}
L.~J. Boya and J.~Casahorran,
\newblock Phys. Lett. B {\bf 215}, 753 (1988); J.\ Phys.\ A {\bf 23},
1645 (1990).
\bibitem{Graham:1998qq}
N.~Graham and R.~L. Jaffe,
\newblock Nucl. Phys. {\bf B544}, 432 (1999). %, hep-th/9808140.
\bibitem{Bordag:2002dg}
M.~Bordag, A.~S. Goldhaber, P.~van Nieuwenhuizen, and D.~Vassilevich,
\newblock {\it Heat kernels and zeta-function regularization for the
mass of the
  susy kink}, hep-th/0203066.
\bibitem{Losev:2000mm}
A.~Losev, M.~A.~Shifman and A.~I.~Vainshtein,
%``Single state supermultiplet in 1+1 dimensions,''
New J.\ Phys.\  {\bf 4}, 21 (2002).
%[arXiv:hep-th/0011027].
%%CITATION = HEP-TH 0011027;%%
\bibitem{rvwprep}
A. Rebhan, P. van Nieuwenhuizen and R. Wimmer, \newblock {\it 
The anomaly in the central charge of the supersymmetric kink from
 dimensional regularization and reduction}, hep-th/0207051.
\bibitem{Uchiyama:1986gf}
A.~Uchiyama,
\newblock Prog. Theor. Phys. {\bf 75}, 1214 (1986).
\bibitem{GSW}M.~B.~Green, J.~H.~Schwarz and E.~Witten, {\it Superstring
theory}, Cambridge Univ. Press, Cambridge, 1987, volume 2,  pp. 279--281.
\bibitem{RJ} R. Jackiw, unpublished.

\end{thebibliography}
\end{document}